\documentclass[11pt,a4paper]{article}
\usepackage{amsmath}
\usepackage{amsfonts}
\usepackage{amssymb}
\usepackage{verbatim}
\usepackage{enumitem}
\usepackage{lineno}
\usepackage[hidelinks]{hyperref}
\usepackage[table]{xcolor}
\usepackage{array}
\usepackage[labelfont=bf,textfont=sf]{caption}
\usepackage{lineno}
\usepackage{microtype}
\usepackage{xfrac}
\usepackage{graphicx}
\usepackage{float}
\usepackage{color}
\usepackage{marvosym}
\usepackage{parskip} 
\usepackage{ulem}
\usepackage[position=top]{subfig}

\usepackage[lmargin=.9in,rmargin=.9in,tmargin=.9in,bmargin=.9in,marginparwidth=100pt]{geometry}

\begin{document}


\newpage

\begin{center}

{\Large \textbf{A comprehensive representation of selection at loci with \\multiple alleles that allows complex forms of genotypic fitness}} 

\bigskip

\small

{\textbf{Nikolas Vellnow$^1$, Toni I. Gossmann$^{1}$ and David Waxman$^{2}$}}

\bigskip
$^1$TU Dortmund University, Computational Systems Biology, \\Faculty of Biochemical and Chemical Engineering,\\ Emil-Figge-Str. 66, 44227 Dortmund, Germany. 
\medskip
\\\Letter~nikolas.vellnow@tu-dortmund.de
\\https://orcid.org/0000-0002-7054-3268
\medskip
\\\Letter~toni.gossmann@tu-dortmund.de
\\https://orcid.org/0000-0001-6609-4116
\bigskip

$^2$Fudan University, Centre for Computational Systems Biology, ISTBI, 
\\220 Handan Road, Shanghai 200433, People's Republic of China.
\medskip
\\\Letter~davidwaxman@fudan.edu.cn
\\https://orcid.org/0000-0001-9093-2108
\end{center}

\bigskip

\normalsize


\begin{abstract}
\sloppy

Genetic diversity is central to the process of evolution. 
Both natural selection and random genetic drift are influenced by the level of genetic diversity of a population; selection acts on diversity while drift samples from it. At a given locus in a diploid population, each individual carries only two alleles, but the population as a whole can possess a much larger number of alleles, with the upper limit constrained by twice the population size. This allows for many possible types of homozygotes and heterozygotes.
Moreover, there are biologically important loci, for example those related to the MHC complex, the ABO blood types, and cystic fibrosis, that exhibit a large number of alleles.
Despite this, much of population genetic theory, and data analysis, are limited to considering biallelic loci. 
However, to the present, what is lacking is a flexible expression for the force of selection that allows an arbitrary number of alleles (and hence an arbitrary number of heterozygotes), along with a variety of forms of fitness.
In this work, we remedy this absence by giving an analytical representation of the force of selection that emphasises the very different roles played by the diversity of the population, and the fitnesses of different genotypes.
The  result presented facilitates our understanding and applies in a variety of different situations involving multiple alleles. This includes situations where fitnesses are: additive, multiplicative, randomly fluctuating, frequency-dependent, and it allows fitnesses which involve explicit gene interactions, such as heterozygote advantage.

\end{abstract}


\clearpage



\clearpage

\section*{Introduction}
\setlength{\parindent}{0pt}

Genetic diversity plays a central role in evolutionary theory since it is the raw material that natural selection can act upon. In population genetics, it is usually measured at the level of single nucleotides or loci, with a focus on biallelic single nucleotide polymorphisms (SNPs) \cite{heyne2023}. However, restricting the analysis of genetic diversity to biallelic loci may lead to biased population genomic inferences, if loci are indeed multiallelic.
In a given diploid individual a biallelic locus can be in one of two homozygous states, or in one heterozygous state. 
However, in the population as a whole, when there are $n$ possible alleles
there can be $n$ different homozygotes and $n(n-1)/2$ different heterozygotes \cite{gillespie2004a, hartl1997}. 
This genetic diversity, which is ultimately created by mutations, may be strongly elevated in mutational hotspots,
 i.e., genomic regions with high mutation rates \cite{nesta2021}, that lead to multiple alleles continually segregating at the same locus. Furthermore, genetic diversity, in the form of multiple alleles, may be both selected for, and maintained by some form of balancing selection, such as heterozygous advantage or frequency-dependent selection \cite{hedrick2012, ayala1974a}. 
 In general, it seems likely that in any population, 
 some loci will have more than two alleles, and hence will have the additional genetic 
 diversity that goes along with this, i.e., more than one possible heterozygote type.

Indeed, there are many empirical examples of multiple alleles occurring at biologically relevant loci. For example, the amylase gene, which is involved in the digestion of starch-rich foods, exhibits multiple copy number variations in dogs, pigs, geladas and humans \cite{axelsson2013, ollivier2016, paudel2013, caldon2024, perry2007}. Other examples, with sometimes \textit{hundreds} of alleles include the major histocompatibility complex (MHC) involved in vertebrate immunity \cite{radwan2020}, the ABO blood types \cite{yamamoto2021}, and single locus genetic diseases like cystic fibrosis \cite{sharma2020}. In general, we may expect loci involved in the interaction between parasites and hosts to have a particularly high allelic richness since the dynamic of host-parasite co-evolution is thought to lead to the maintenance of many alleles via balancing selection \cite{ebert2020, cornetti2024}. Furthermore, the self-incompatibility locus in flowering plants is known to have an extremely high allelic richness \cite{takayama2005}, a phenomenon that is thought to be caused by strong negative frequency-dependent selection \cite{wright1939, yokoyama1979, charlesworth2005}.

Recent methodological advances in the field of genome sequencing and genome analysis have made the consideration of multiple alleles much more practical. In order to reliably detect alleles segregating at low frequencies (e.g., $<0.05$) it is necessary to sequence at very high depth \cite{gossmann2022}. Performing \textit{whole genome sequencing} at such a high sequencing depth used to be prohibitively expensive but is now affordable by many researchers. Such deep sequencing will make the detection of rare alleles, and therefore the study of loci with multiple alleles, possible. We also expect that many loci, previously thought to be biallelic, will be discovered to be multiallelic when subject to deep sequencing. However, population genetic theory mostly focuses on biallelic loci, e.g. when using Watterson's $\theta$ or Tajima's $D$ \cite{watterson1975, tajima1989}, and many genomics tools, and the associated software, assume (as indicated by their default settings) the presence of only two alleles \cite{campbell2016}. Therefore, focusing on population genetic studies in scenarios with more than two alleles is clearly needed.

Here, we present a framework for the force of selection
that can incorporate an arbitrary number of alleles, different forms of genotypic 
fitness, and which clearly differentiates the different roles played by genetic diversity and genotypic fitness. This framework allows the inclusion of interesting selection regimes, e.g. frequency-dependent selection, and gives insights into how the interplay between diversity and fitness is influenced by such selection regimes.

\section*{Motivation from a simple model}

A very basic form of diversity is the genetic variation which exists at a single locus, and in what follows we shall often use \textit{diversity} and \textit{variation} as synonymous terms. 
Generally, natural selection acts on (biallelic or multiallelic) variation within a population and generates changes within the population. In addition to this, random genetic drift samples amongst existing variation within a population, and produces random changes. Thus both natural selection and random genetic drift only operate when there is variation within a population. This basic observation suggests that there may be a connection between the mathematical representation of these two distinct evolutionary `forces' - selection and drift, which both vanish in the absence of variation. In this work we shall establish this connection by co-opting an expression involving population variability, that appears in formulae for random genetic drift, to obtain a compact formula for the selective force when there are multiple alleles.

For the simplest model involving a locus with two alleles, the connection between selection and drift can be readily seen, as we now show.

Consider a randomly mating diploid population with discrete generations, where
a single biallelic locus is subject to multiplicative selection. Let $x$ denote
the frequency of a focal allele at the locus in a given generation. By the
beginning of the next generation we have \cite{gillespie2004a} that:

\begin{enumerate}
[label=(\roman*)]

\item selection contributes the amount $sx(1-x)$ to the change in frequency,
where $s$ is a selection coefficient associated with  the
focal allele (we have assumed $|s| \ll1$), 

\item random genetic drift contributes an amount $x(1-x)/(2N)$ to the variance of
the frequency, where $N$ is the population size\footnote{We work within the
framework of ideal Wright-Fisher model, which depends on the census population size, $N$, 
and not the effective population size, $N_{e}$. However, incorporation of $N_e$ into 
this model is possible \cite{zhao2016}.}.
\end{enumerate}
We note that these effects, from selection and drift, both contain an
\textit{identical function of frequency}, namely $x(1-x)$, which vanishes in
the absence of variation (at $x=0$ and $x=1$). This shows a common dependency of
these evolutionary forces on a basic measure of 
diversity of the population at the locus, namely $x(1-x)$.

We use the case above with two alleles to motivate a question that leads to a formula 
for selection at a locus with multiple alleles:

\begin{quote}
\textit{When there are more than two alleles at a locus, do the
effects of natural selection and random genetic drift have a common
dependence on the diversity of the population?}
\end{quote}

To answer this question, we first establish the model that we shall use.

\section*{Model}

We adopt a Wright-Fisher model for a population of diploid individuals in which there 
is an equal sex ratio and no sexual dimorphism. Generations are discrete, and labelled 
by $t=0,1,2,\ldots$. At the start of a generation there are $N$ adult individuals in 
the population. The life cycle is:

\medskip

\noindent (i) reproduction of adults by random mating, with all adults making the same 
very large contribution to the number of offspring, independent of parental genotype;\newline  
\noindent (ii) death of all adults; \newline  
\noindent (iii) offspring subject to selection at a single unlinked locus 
with $n$ different alleles; \newline  
\noindent (iv) non-selective thinning of the population to $N$ individuals - who constitute the 
adults of the next generation.

\medskip

There is no restriction on the number of alleles at the locus, so that $n$ can have 
the values $2$, $3$, $4$, $\ldots$, and we label the $j$'th allele as $B_{j}$, with $j=1,2,...,n$.

To describe the population, we use an $n$ component column vector,
$\mathbf{X}(t)$, whose $j$'th element, $X_{j}(t)$, is the frequency of allele
$B_{j}$ in adults, at the start of generation $t$, immediately prior to reproduction.

We take the $B_{i}B_{j}$ genotype to have a fitness proportional to
$1+A_{i,j}$, i.e.,
\begin{equation}
\text{fitness of }B_{i}B_{j}\text{ genotype}\quad\propto\quad1+A_{i,j}.\label{fitness}
\end{equation}
The quantity $A_{i,j}$ is a measure of the selection acting on the $B_{i}
B_{j}$ genotype\footnote{The $A_{i,j}$ may depend on allele
frequencies, in which case selection is \textit{frequency dependent}, and in
generation $t$ the $A_{i,j}$ will depend on $\mathbf{X}(t)$. In what follows
we shall treat the $A_{i,j}$ as constants, unless we state otherwise.}, and could be described as the
`fitness effect' of this genotype

In what follows we shall use $\mathbf{A}$ to denote the $n\times n$ matrix
whose $(i,j)$ element is $A_{i,j}$. Assuming the absence of genomic imprinting
or other epigenetic phenomena \cite{peters2014} we have $A_{i,j}=A_{j,i}$, i.e., the
matrix $\mathbf{A}$ is \textit{symmetric}.

\subsection*{Notation}

For convenience we present here the remainder of the notation that we will use
in the rest of this work.

Generally, the vector of allele frequencies at any time $t$, namely
$\mathbf{X}(t)$, is a random variable with the properties of being
non-negative and normalised to unity in the sense that $X_{j}(t)\geq0$ and $\sum
_{j=1}^{n}X_{j}(t)=1$. We write realisations (or non-random examples) of
$\mathbf{X}(t)$ as $\mathbf{x}$ and these are also non-negative and normalised to unity.

\bigskip

\noindent With a $T$ superscript denoting the transpose of a matrix, we use the
notation where:

\begin{itemize}

\item $\delta_{i,j}$ denotes a Kronecker delta, which takes the value $1$
when the  indices $i$ and $j$ are equal, and is zero otherwise 

\item $\mathbf{F}$ denotes an $n$ component column vector with all elements
$1$:
\begin{equation}
\mathbf{F}=(1,1,\ldots,1)^{T}\label{F def}
\end{equation}

\item $\mathbf{I}$ denotes the $n\times n$ identity matrix 

\item $\mathbf{V}(\mathbf{x})$ denotes an $n\times n$ matrix which depends
on the vector $\mathbf{x}$,  and has elements given by
\begin{equation}
V_{i,j}(\mathbf{x})=x_{i}\delta_{i,j}-x_{i}x_{j}\qquad i,j=1,2,\ldots,n.\label{Vij}
\end{equation}

\end{itemize}

\section*{Dynamics of the model}

The frequency of \textit{all} alleles in generation $t+1$, i.e., after reproduction,
selection and number thinning have occurred in generation $t$, can be written
in the form of the single equation
\begin{equation}
\mathbf{X}(t+1)=\mathbf{X}(t)+\mathbf{D}(\mathbf{X}(t))+\boldsymbol{\xi}(t).
\label{discrete t dyn}
\end{equation}
where $\mathbf{X}(t)$, $\mathbf{D}(\mathbf{X}(t))$ and $\boldsymbol{\xi}(t)$
are all $n$ component column vectors. The quantities $\mathbf{D}
(\mathbf{X}(t))$ and $\boldsymbol{\xi}(t)$ can be called `evolutionary forces'
since they both change $\mathbf{X}(t)$. They arise from natural
selection and random genetic drift, respectively.

To describe Eq. (\ref{discrete t dyn}) in greater detail, we begin with the
force of genetic drift, namely the vector $\boldsymbol{\xi}(t)$, with
elements $\xi_{j}(t)$. For any realisation of $\mathbf{X}(t)$, the expected
value of $\boldsymbol{\xi}(t)$ over replicate populations vanishes, and we
write this as
$E\left[  \boldsymbol{\xi}(t)|\mathbf{X}(t)=\mathbf{x}\right]  =\mathbf{0}$.

The variance-covariance matrix of $\boldsymbol{\xi}(t)$, conditional on
$\mathbf{X}(t)=\mathbf{x}$, is given approximately\footnote{The
approximation neglects the effect of selection 
on the right hand side of Eq. (\ref{cov=V/2N}) \cite{ewens2004}.} by
\begin{equation}
E[\boldsymbol{\xi}(t)\boldsymbol{\xi}^{T}(t)|\mathbf{X}(t)=\mathbf{x}
]\simeq\frac{\mathbf{V}(\mathbf{x})}{2N} \label{cov=V/2N}
\end{equation}
where the matrix $\mathbf{V}(\mathbf{x})$ of Eq. (\ref{Vij}) enters - see
Appendix A for details. 
Thus the $(i,j)$ element of the variance-covariance
matrix of $\boldsymbol{\xi}(t)$ is given by $E[\xi_{i}(t)\xi_{j}
(t)|\mathbf{X}(t)=\mathbf{x}]\simeq V_{i,j}(\mathbf{x})/(2N)$\textbf{. }

When there are multiple alleles, the matrix $\mathbf{V}(\mathbf{x})$ in Eq.
(\ref{cov=V/2N}) explicitly exposes the dependence of genetic drift on
the diversity of the population. 
The only time $\mathbf{V}(\mathbf{x})$ vanishes
(has all elements zero) is when a single allele is at a frequency of unity
(and all other alleles are at zero frequency).

Consider now the selective force in Eq. (\ref{discrete t dyn}), namely
$\mathbf{D}(\mathbf{X}(t))$. There are different ways to write $\mathbf{D}
(\mathbf{x})$ but in Appendix B we show that it can be compactly written as
\begin{equation}
\mathbf{D}(\mathbf{x})=\frac{\mathbf{V}(\mathbf{x})\mathbf{Ax}}{1+\mathbf{x}
^{T}\mathbf{Ax}}\qquad\text{general case.} \label{D(x)}
\end{equation}
In this representation of $\mathbf{D}(\mathbf{x})$ we see the presence of
the matrix $\mathbf{A}$, which is associated with genotypic fitness - see Eq.
(\ref{fitness}). We also see the matrix $\mathbf{V}(\mathbf{x})$, which naturally
emerged as the feature of random genetic drift that exposed its dependence on
the population's diversity. 
Evidently, the matrix $\mathbf{V}(\mathbf{x})$
also plays a key role in the form of the evolutionary force of selection. This
connection is even clearer under the continuous time, continuous state,
diffusion approximation of Eq. (\ref{discrete t dyn}),
which says 
that over a small time interval $dt$, the change in
frequencies, $d\mathbf{X}$, arises from the sum of effects
of selective and drift forces. With $\mathbf{D}
(\mathbf{x})\simeq\mathbf{V}(\mathbf{x})\mathbf{Ax}$ and omitting $t$
arguments, the diffusion approximation of Eq. (\ref{discrete t dyn}) takes the form
\begin{equation}
d\mathbf{X}=\mathbf{V}(\mathbf{X})\mathbf{AX}dt+\sqrt{\frac{\mathbf{V}
(\mathbf{X})}{2N_{e}}}d\mathbf{W} \label{diffusion}
\end{equation}
where we have introduced the effective population size, $N_{e}$, and
also the column vector $\mathbf{W}\equiv\mathbf{W}(t)$ that has random elements, 
representing the stochasticity of genetic drift. Thus the same
frequency-dependent matrix, $\mathbf{V}(\mathbf{X})$ appears in the terms
arising from selection and drift in Eq. (\ref{diffusion}) and it is natural to
generally associate this matrix with a description of 
\textit{diversity} of the population.

With Eq. (\ref{D(x)}) or Eq. (\ref{diffusion}), we have an answer to the question
in the Introduction:

\begin{quote}
\textit{When there are multiple alleles, selection and drift have a common
dependence on the 
diversity of the population which, when allele frequencies are given by $\mathbf{x}$, is described by $\mathbf{V}(\mathbf{x})$}.
\end{quote}

\newpage

\section*{Special cases of $\mathbf{D}(\mathbf{x})$}

We next explore some special cases of the selective force given in Eq.
(\ref{D(x)}).

\subsection*{Additive selection}

Under additive selection, the $B_{i}B_{j}$ genotype has a fitness proportional
to $1+A_{i,j}=1+s_{i}+s_{j}$ where $s_{i}$ is a selection coefficient
associated with allele $B_{i}$. With $\mathbf{s}$ an $n$ component column
vector whose $i$'th element is $\mathbf{s}_{i}$, we have
\begin{equation}
\mathbf{s}=(s_{1},s_{2},\ldots,s_{n})^{T} \label{s}
\end{equation}
and under additive selection we can write the matrix $\mathbf{A}$ as
\begin{equation}
\mathbf{A}=\mathbf{sF}^{T}+\mathbf{Fs}^{T} \label{A_additive}
\end{equation}
because its elements are precisely given by $A_{i,j}=s_i+s_j$.

Given that $\mathbf{x}$ represents a column vector of possible allele
frequencies, it has the property $\sum_{j=1}^{n}x_{j}=1$, which can be
written as $\mathbf{F}^{T}\mathbf{x}=1$, and this leads to $\mathbf{V}
(\mathbf{x})\mathbf{F}$ being a vector with all elements zero (using Eq. (\ref{Vij})). Thus
\begin{equation}
\mathbf{V}(\mathbf{x})\mathbf{Ax}=\mathbf{V}(\mathbf{x})\left(  \mathbf{sF}
^{T}+\mathbf{Fs}^{T}\right)  \mathbf{x}=\mathbf{V}(\mathbf{x})\mathbf{s}.
\end{equation}
Also we have
\begin{equation}
1+\mathbf{x}^{T}\mathbf{Ax}\mathbf{=}1+\mathbf{x}^{T}\left(  \mathbf{sF}
^{T}+\mathbf{Fs}^{T}\right)  \mathbf{x}=1+2\mathbf{s}^{T}\mathbf{x}.
\end{equation}
These lead to
\begin{equation}
\mathbf{D}(\mathbf{x})=\frac{\mathbf{V}(\mathbf{x})\mathbf{s}}{1+2\mathbf{s}
^{T}\mathbf{x}}\qquad\text{additive selection.} \label{D add}
\end{equation}
In the special case of $n=2$ alleles, with $s_{1}=s$, $s_{2}=0$, and writing 
$x=x_{1}$, we obtain a selective force on the frequency of allele $B_{1}$ of
$D_{1}(\mathbf{x})=sx(1-x)/\left(  1+2sx\right)  $.

\subsection*{Multiplicative selection}

Under multiplicative selection, the $B_{i}B_{j}$ genotype has a fitness
proportional to $1+A_{i,j}=\left(1+s_{i}\right)\left(1+s_{j}\right)$
and again we shall refer to $s_{i}$ as the selection coefficient associated
with allele $B_{i}$. In this case, again using Eq. (\ref{s}), we can write
\begin{equation}
\mathbf{A}=\left(  \mathbf{F}+\mathbf{s}\right)  \left(  \mathbf{F}
+\mathbf{s}\right)  ^{T}-\mathbf{FF}^{T}.
\end{equation}
We then have $\mathbf{V}(\mathbf{x})\mathbf{Ax=V}(\mathbf{x})\mathbf{s}\left(
1+\mathbf{s}^{T}\mathbf{x}\right)  $ and $1+\mathbf{x}^{T}\mathbf{Ax}=\left(
1+\mathbf{s}^{T}\mathbf{x}\right)  ^{2}$ and hence
\begin{equation}
\mathbf{D}(\mathbf{x})=\frac{\mathbf{V}(\mathbf{x})\mathbf{s}}{1+\mathbf{s}
^{T}\mathbf{x}}\qquad\text{multiplicative selection.} \label{D mult}
\end{equation}
In the special case of $n=2$ alleles, with $s_{1}=s$, $s_{2}=0$, and writing $x=x_{1}$, 
we obtain a selective force on the frequency of allele $B_{1}$ of
$D_{1}(\mathbf{x})=sx(1-x)/\left(  1+sx\right)  $.

\subsection*{Heterozygote advantage}

We next consider one of the simplest forms of selection that acts genuinely at
the level of genotypes, and cannot be devolved to properties of individual
alleles. This is selection where all heterozygotes have a common fitness
advantage of $\sigma$ relative to all homozygotes. That is
\begin{equation}
A_{i,j}=\left\{
\begin{array}
[c]{ccc}
0, &  & i=j\\
&  & \\
\sigma, &  & i\neq j.
\end{array}
\right.
\end{equation}
This corresponds to
\begin{equation}
\mathbf{A}=\sigma\left(  \mathbf{FF}^{T}-\mathbf{I}\right)  .
\end{equation}
Using this result in Eq. (\ref{D(x)}) yields
\begin{equation}
\mathbf{D}(\mathbf{x})=-\frac{\sigma\mathbf{V}(\mathbf{x})\mathbf{x}}
{1+\sigma\left(  1-\mathbf{\mathbf{x}^{T}\mathbf{x}}\right)  }\qquad
\text{heterozygote selection.}   \label{heterozygote selection}
\end{equation}
In the special case of $n=2$ alleles, and writing $x=x_{1}$, we obtain a
selective force on the frequency of allele $B_{1}$ of $D_{1}(\mathbf{x}
)=\sigma x(1-x)(1-2x)/\left[  1+2\sigma x(1-x)\right]  $.

\subsection*{Fluctuating selection}

We consider the situation where fitnesses randomly change \cite{huerta-sanchez2008, gossmann2014a}, from one generation
to the next, due, for example, to random environmental effects on the population. Here we
give one example based on Eq. (\ref{D(x)}). Assuming fitnesses are not
correlated over time, it is then appropriate to directly average over the
fluctuating fitnesses. The standard way of proceeding, \textit{as far as the selective force is 
concerned} (we do not consider the change in the variance of allelic effects), is to first expand 
the selective force to second order in selection coefficients, and then
average the result \cite{jensen1969}. Here, we shall interpret 
the analogue of selection coefficients as the elements of the matrix $\mathbf{A}$, that appears in Eq. (\ref{D(x)}),
and we will expand $\mathbf{D}(\mathbf{x})$ to second order in $\mathbf{A}$.
This leads to $\mathbf{D}(\mathbf{x})\simeq\mathbf{V}(\mathbf{x}
)\mathbf{Ax}-\mathbf{V}(\mathbf{x})\mathbf{Ax}\left(  \mathbf{x}^{T}
\mathbf{Ax}\right) $.

As an illustrative example, we take
\begin{equation}
A_{i,j}=\sqrt{\lambda}(Z_{i,j}+Z_{j,i})/2\label{A=Z+Z}
\end{equation}
where the $Z_{i,j}$ are, for all $i$ and $j$, independent normal random
variables with mean zero and variance one, while $\lambda$ is a measure of variance of
fitness effects.
The form of $A_{i,j}$ adopted in
Eq. (\ref{A=Z+Z}) ensures that $A_{i,j}=A_{j,i}$. It also leads to the mean
value of $\mathbf{A}$ vanishing (the average effect of selection on any
genotype is zero) which we write as $\mathbf{\bar{A}}=\mathbf{0}$.
Furthermore with $A_{i,j}$ given by Eq. (\ref{A=Z+Z}), to calculate the average of 
$\mathbf{D}(\mathbf{x})$, we require knowledge of the average of
$A_{i,j}A_{k,l}$ for arbitrary $i$, $j$, $k$ and $l$. This is given by 
\begin{equation}
\overline{A_{i,j}A_{k,l}} = \lambda\left( \delta_{i,k}\delta_{j,l}
+\delta_{i,l}\delta_{j,k}\right) /2
\end{equation}
and using this result leads to\footnote{The model adopted for fluctuating selection,
where $A_{i,j}$ is given by Eq. (\ref{A=Z+Z}), leads to a simple result for
the averaged force of selection, in terms of one parameter, $\lambda$, along
with the matrix $\mathbf{V}(\mathbf{x})$ and vector $\mathbf{x}$ - both in
unaltered form. However, the model has the feature that the variance of
fitness effects of homozygotes is \textit{twice} that of heterozygotes
($\overline{A^{2}_{i,i}}=\lambda$, but $\overline{A^{2}_{i,j}}=\lambda/2$ for
$i \neq j$).}
\begin{equation}
\overline{\mathbf{D}}(\mathbf{x})\simeq-\overline{\mathbf{V}(\mathbf{x}
)\mathbf{Ax}\left(  \mathbf{x}^{T}\mathbf{Ax}\right)  }=-\lambda \left(\mathbf{x}^{T}\mathbf{x}\right)
\mathbf{V}(\mathbf{x})\mathbf{x}.     \label{fluctuating}
\end{equation}
In the special case of $n=2$ alleles, and writing $x=x_{1}$, we obtain an
averaged selective force on allele $B_{1}$ of $\overline{D}_{1}(\mathbf{x}
)=\lambda x(1-x)(1-2x)\left[x^{2}+(1-x)^{2}\right]$.

\subsection*{Frequency dependent selection}

We can model negative frequency-dependent selection by 
allowing the fitness of a genotype to depend on the genotype's frequency, such that the fitness is a \textit{decreasing function} of genotype frequency.
This results in the matrix of fitness effects, $\mathbf{A}$, becoming $\mathbf{A(x)}$, 
where $\mathbf{x}$ can be thought of as the vector of current allele frequencies. The $B_{i}B_{j}$ genotype now has a fitness 
proportional to $1 + A_{i,j}(\mathbf{x})$
and we take 
\begin{equation}
A_{i,j}(\mathbf{x}) = - c x_i x_j  \label{fitness_nfds}
\end{equation}
where $c$ is a constant\footnote{A more general form of negative frequency-dependent 
selection is given by $A_{i,j}(\mathbf{x}) = - c_{i,j} x_i x_j$ with $c_{i,j}=c_{j,i}$ and $0 \leq c_{i,j} < 1$.} that lies in the range $0 \leq c < 1$.

Since we can write $\mathbf{A(x)=}-c\mathbf{xx}^{T}$, the form of the selective force 
following from Eq. (\ref{D(x)})
is
\begin{equation}
\mathbf{D}(\mathbf{x})=-\frac{c\left(  \mathbf{x}^{T}\mathbf{x}\right)
}{1-c\left(  \mathbf{x}^{T}\mathbf{x}\right)  ^{2}}\mathbf{V}(\mathbf{x}
)\mathbf{x}.\label{freqsel}
\end{equation}

In the special case of $n=2$ alleles, and writing $x=x_{1}$, we obtain a selective
force on the frequency of allele $B_{1}$ of $D_{1}(\mathbf{x})
=cx(1-x)(1-2x)\frac{\left[x^{2}+(1-x)^{2}\right]}{1-c\left[x^{2}+(1-x)^{2}\right]^2}$.

\section*{Illustrations of the dynamics}
To illustrate the dynamics that naturally arises from our framework in some interesting cases, we calculated allele frequency trajectories numerically. The detailed scripts are available as a github repository (\url{https://github.com/NikolasVellnow/selection\_multiallelic}). 
In particular, for a locus with $n=3$ alleles, 
we calculated three pairs of frequency trajectories, for a number of different scenarios,
over $5 \times 10^3$ generations. 
In each pair of trajectories, one was deterministic and the other stochastic, corresponding to population 
sizes of infinity and $10^3$, respectively.

 In the results we present, the finite (but large) number of generations
we consider has the consequence that the deterministic trajectories closely approach, but never 
actually reach the end point they would ultimately achieve. In what follows, we shall make statements 
about `achieving fixation' or `achieving a stable polymorphism' that should be interpreted to
hold to an accuracy of machine precision ($1$ part in  $10^{14}$).

Furthermore, to aid visualisation of the trajectories given in the figures of this section, 
we have included a coloured planar triangular region, that all trajectories are confined to lie in 
(as a consequence of their three allele frequencies summing to unity at all times).

\begin{figure}[H]
    \includegraphics[width=1.\textwidth]{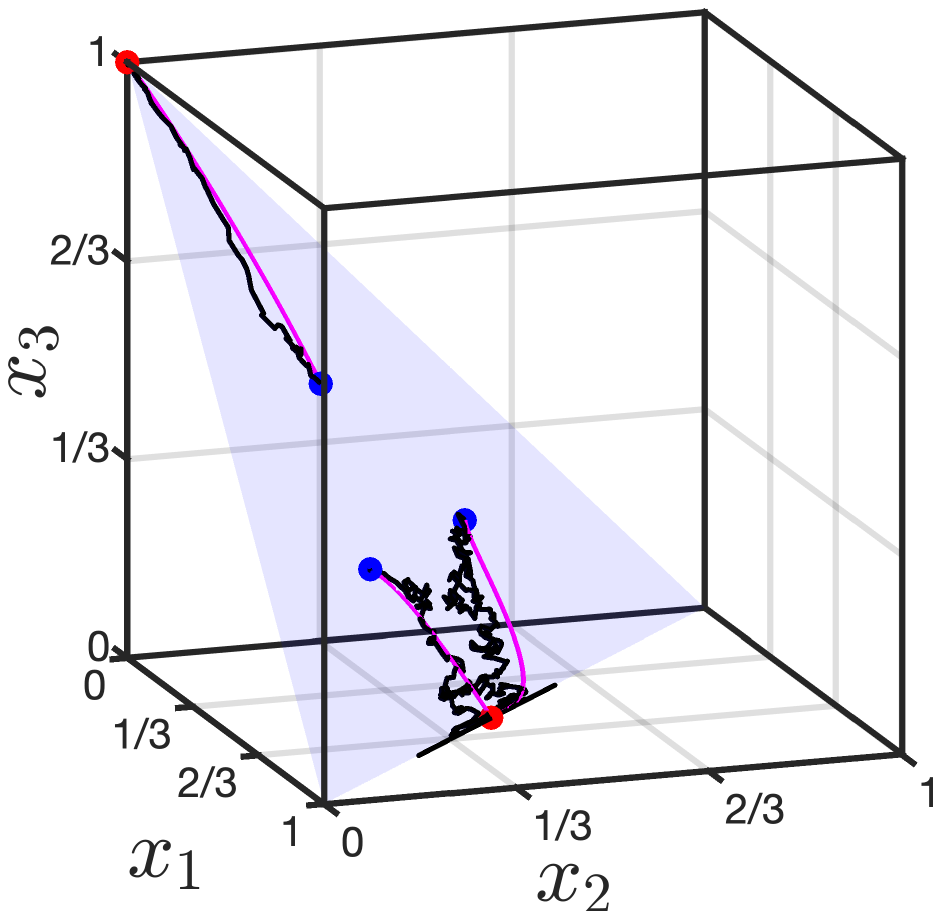}
    \centering
    \caption{\textbf{Allele frequency trajectories 
    for the case of a single randomly generated set of fitness effects} for $n=3$ alleles, with frequencies $x_1$, $x_2$ and $x_3$. Three pairs of trajectories are plotted, each starting from a common point (blue dot). In each pair, one trajectory is deterministic (magenta) and is appropriate to an infinite population, while the other is stochastic (black), and is appropriate to a population of size $N = 10^3$. The trajectories were run for a total of $5 \times 10^3$ generations. End points of deterministic trajectories are given by red dots.} \label{fig:det_stoch_random}
\end{figure}

For Figure \ref{fig:det_stoch_random}, we proceeded by first calculating the dynamics with a randomly generated
set of fitness effects, based on a normal distribution, as given in Eq. (\ref{A=Z+Z}). The set of fitness values was generated just once, at the beginning of the simulation, and was then used throughout.
We simulated a scenario where three populations had the starting allele frequencies $(0.50, 0.25, 0.25)$, $(0.25, 0.50, 0.25)$, and $(0.25, 0.25, 0.50)$. We observe that the deterministic end point of the three populations may depend on the initial frequencies (see Figure \ref{fig:det_stoch_random}). The populations with starting frequencies $(0.50, 0.25, 0.25)$, $(0.25, 0.50, 0.25)$ both reached the end point
$(0.5595, 0.4405, 0)$.
This end point, under deterministic dynamics, is a stable polymorphism, where alleles $B_1$ and $B_2$ are maintained in the population while allele $B_3$ is lost. By contrast, the population with the starting frequencies  $(0.25, 0.25, 0.50)$ 
reached the end point $(0, 0, 1)$, corresponding to fixation of allele $B_3$.

\bigskip

\begin{figure}[H]
    \includegraphics[width=\textwidth]{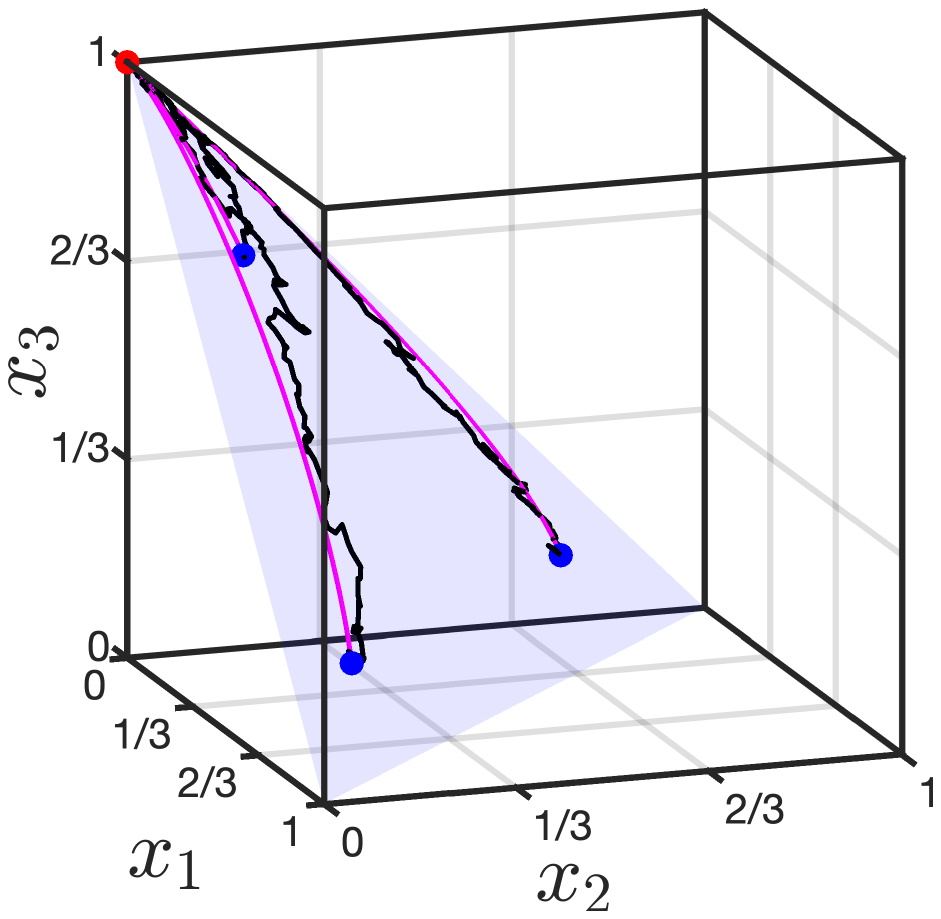}
    \centering
    \caption{\textbf{Allele frequency trajectories for the case of additive fitness}, for $n=3$ alleles, with frequencies $x_1$, $x_2$ and $x_3$. Three pairs of trajectories are plotted, each starting from a common point (blue dot). In each pair, one trajectory is deterministic (magenta) and is appropriate to an infinite population, while the other is stochastic (black), and is appropriate to a population of size $N=10^3$. The trajectories were run for a total of $5 \times 10^3$ generations, with the common end point of all deterministic trajectories given by the red dot.}
    \label{fig:det_stoch_additive}
\end{figure}

In Figure \ref{fig:det_stoch_additive} we illustrate the dynamics when fitnesses are  additive  
(Eq. \ref{A_additive}). For this, we simulated a scenario where three populations had the starting allele 
frequencies $(0.70, 0.15, 0.15)$, $(0.15, 0.07, 0.15)$, and $(0.15, 0.15, 0.70)$. The set of fitness 
values was calculated from the allelic selection coefficients $(0.01, 0.03, 0.06)$ which were then used throughout. In this case, whether following deterministic or stochastic trajectories, allele $B_{3}$ reached fixation in all three pairs of populations, independent of the starting frequencies (Figure \ref{fig:det_stoch_additive}).

\begin{figure}[H]
    \includegraphics[width=\textwidth]{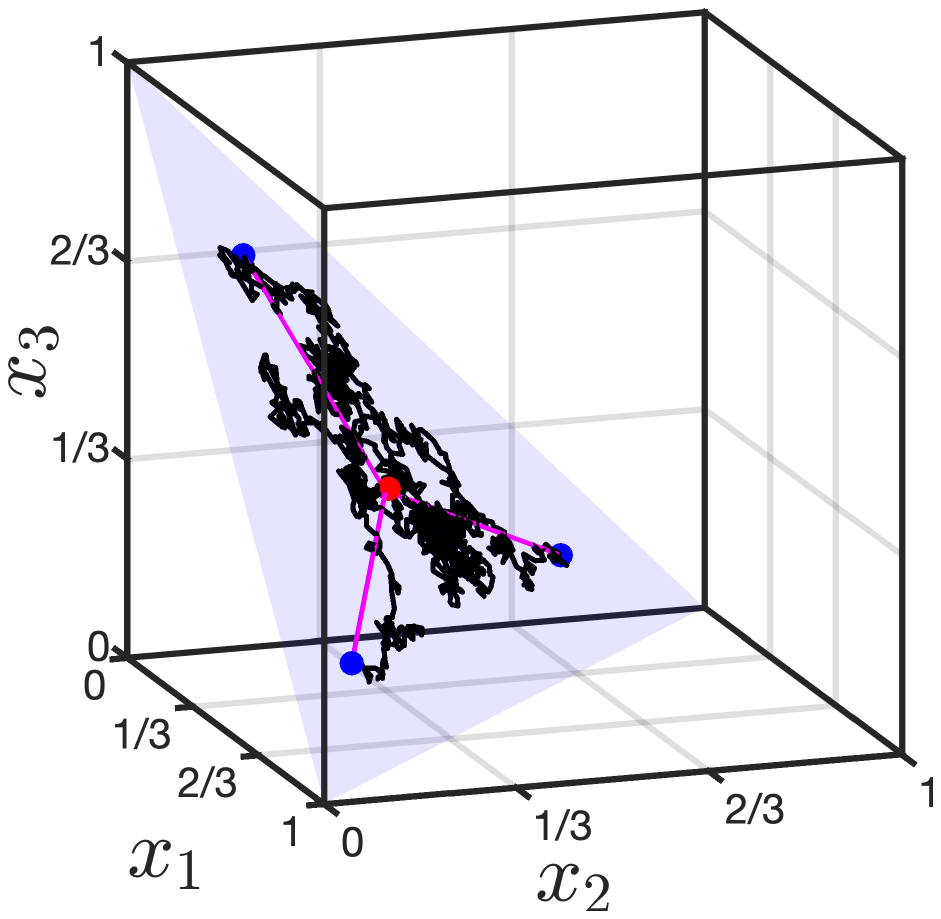}
    \centering
    \caption{\textbf{Allele frequency trajectories for the case of negative frequency-dependent selection} for $n=3$ alleles, with frequencies $x_1$, $x_2$ and $x_3$. Three pairs of trajectories are plotted, each starting from a common point (blue dot). In each pair, one trajectory is deterministic (magenta) and is appropriate to an infinite population, while the other is stochastic (black), and is appropriate to a population of size $N=10^3$. The trajectories were run for a total of $5 \times 10^3$ generations, and the common end point of all deterministic trajectories is given by the red dot.}
    \label{fig:det_stoch_nfds}
\end{figure}

 In Figure \ref{fig:det_stoch_nfds} we give an example of a type of balancing selection. We calculated the dynamics for frequency-dependent selection (Eq. \ref{fitness_nfds}). We simulated a scenario with three populations, that had the starting allele frequencies $(0.70, 0.15, 0.15)$, $(0.15, 0.70, 0.15)$, and $(0.15, 0.15, 0.70)$, where the strength of frequency-dependent selection was $c = 0.05$. The set of fitness values was updated every generation, based on the current allele frequencies (cf. Eq. (\ref{fitness_nfds})). Here, all three populations, following a deterministic trajectory, reached the same stable 
 polymorphism, that had the intermediate end point frequencies $(1/3, 1/3, 1/3)$, independent of the starting frequencies (see Figure \ref{fig:det_stoch_nfds}). All populations following a stochastic trajectory drifted around the point $(1/3, 1/3, 1/3)$ and, over the time of simulation, exhibited a polymorphism that included  all alleles (Figure \ref{fig:det_stoch_nfds}) 
 
To calculate diverse cases of fitness, it is only necessary to change the form of the matrix $\mathbf{A(x)}$, 
and decide if it is calculated only once, at the beginning of the simulation, or whether it has to be updated after each generation (e.g., as in the case of frequency-dependent selection). This highlights the flexibility with which our framework can implement different forms of selection on multiallelic loci.

\section*{Discussion}

In this work we have presented a generalised framework to model multiallelic genetic diversity that incorporates complex forms of selection. We provide detailed scripts as supplemental material to this manuscript, which allow interested researchers to tailor this powerful population genetic framework to their specific research questions. Our approach not only allows the incorporation of complex forms of selection, that act at the level of genotypes, but most importantly, it may be applied 
when there are an arbitrary number of alleles. Therefore, unlike many treatments that assume biallelic loci, our approach can incorporate additional alleles that may be 
at (very) low frequencies. This is in particular relevant when: (i) the genetic diversity of a single locus is very high and many alleles exist \cite{radwan2020, yamamoto2021, sharma2020} or (ii) the effective population size at a particular locus is very high \cite{gossmann2011, smith1974, charlesworth1993}.

The comprehensive form for the deterministic force of selection, 
that we have presented in this work, can be integrated into a broader framework 
involving stochastic variation. 
It is possible that in a multiallelic sample (many) alleles segregate at low frequencies and for these alleles stochastic effects play a pronounced role. Rare alleles are more likely to be lost, due to random fluctuations in allele frequencies, than relatively common alleles. With multiple alleles, allele frequency changes due to drift can also result in complex fluctuations. Since there are more alleles, the overall variance in allele frequencies across generations can be higher than in a simple two-allele system. In loci with many alleles, genetic drift should gradually reduce the overall genetic diversity over many generations \cite{kimura1955, latter1969}. However, in larger populations or populations with balancing selection (such as heterozygote advantage or frequency-dependent selection), genetic diversity may be maintained despite the random effects of drift \cite{kimura1964, lewontin1978, spencer1988}. Discovering alleles that may contribute to such processes is dependent on correct identification of the total genetic variation at a given genetic locus. Our approach can help in the understanding of the tight interplay between increased mutations, drift and the action of balancing selection. An example, where only certain types of allelic combinations 
lead to a heterosis effect, could be straightforwardly modelled within our framework.

Since our model allows the incorporation of complex forms of selection, which act at the level of the
genotype, it can be used to study a wide range of questions. For this, it is only necessary to define
the matrix of fitness effects, $\mathbf{A}(\mathbf{x})$, in accordance with the specific research
question. For instance, we think that the interesting model used by \cite{haasl2013}, in which
they considered genotypic fitness to be dependent on the length of the microsatellite allele at a 
focal locus \cite{haasl2013, haasl2014}, could also be modelled as a special case of our framework.
Furthermore, arbitrarily complex forms of incomplete dominance could also be modelled.

There are some direct spin-offs from the general framework that we have presented. 
For example, it exposes relations between the selective forces arising 
from different forms of fitness. Consider the selective forces in Eqs. (\ref{heterozygote selection}), (\ref{fluctuating}), and (\ref{freqsel}), that arise, respectively, 
from fitnesses that act on heterozygotes, or are fluctuating, or that lead to negative frequency-dependent selection. While all these forms of fitness are different,
the selective forces that arise all have very similar forms. In particular all three 
expressions for the force are of the form $-\kappa\mathbf{V}(\mathbf{x})\mathbf{x}$, where $\kappa$
depends on $\mathbf{x}$ but has a purely numerical value (it is not a vector or matrix). 
This means all three  selective forces push 
allele frequencies in precisely the same `direction', but by different amounts, because the coefficient, $\kappa$, generally has different forms and different values in the three cases.

\begin{figure}[H]
    \centering
    \subfloat[\raggedright]{{\includegraphics[width=0.8\textwidth]{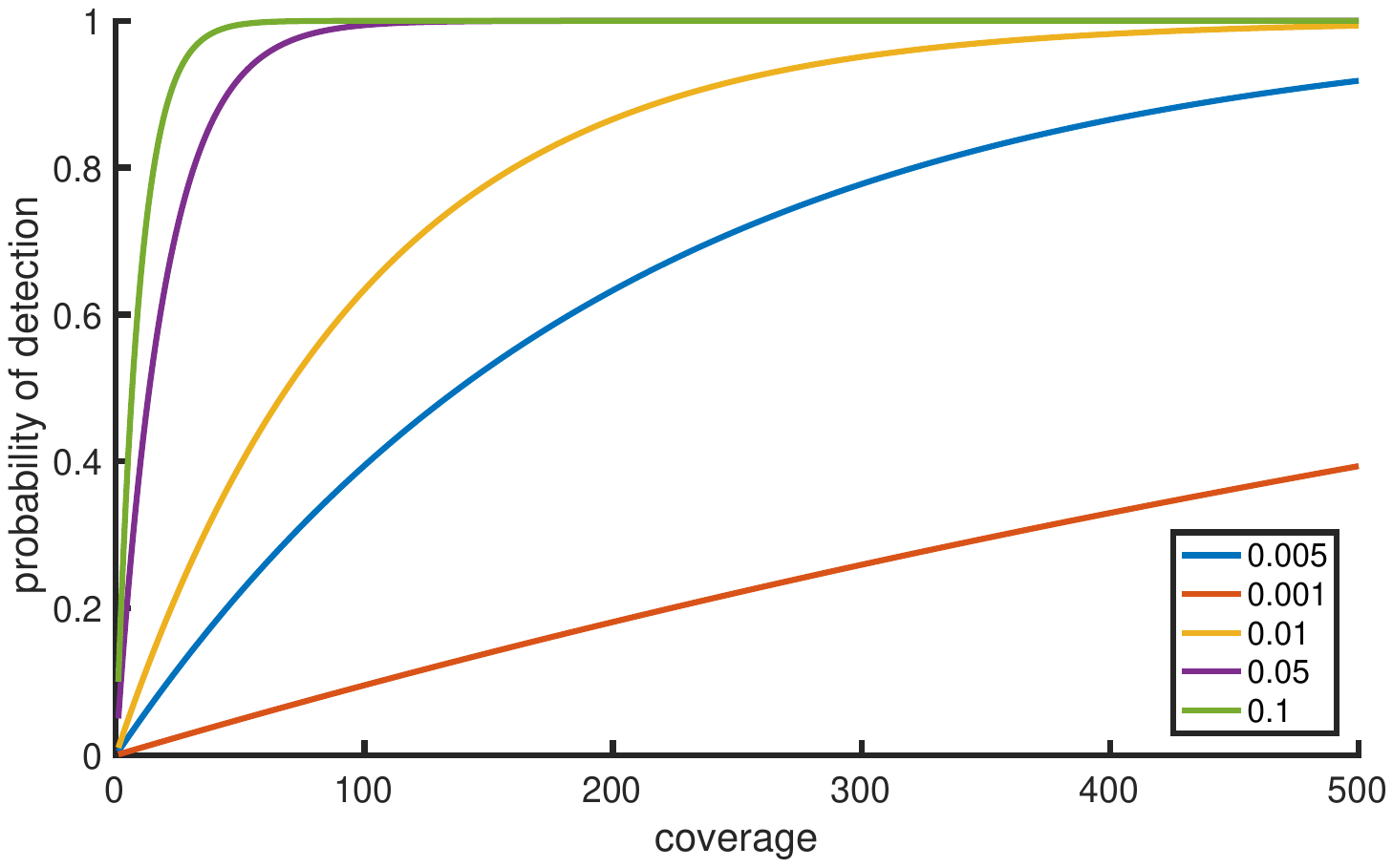} }}
    \qquad
    \subfloat[\raggedright]{{\includegraphics[width=0.8\textwidth]{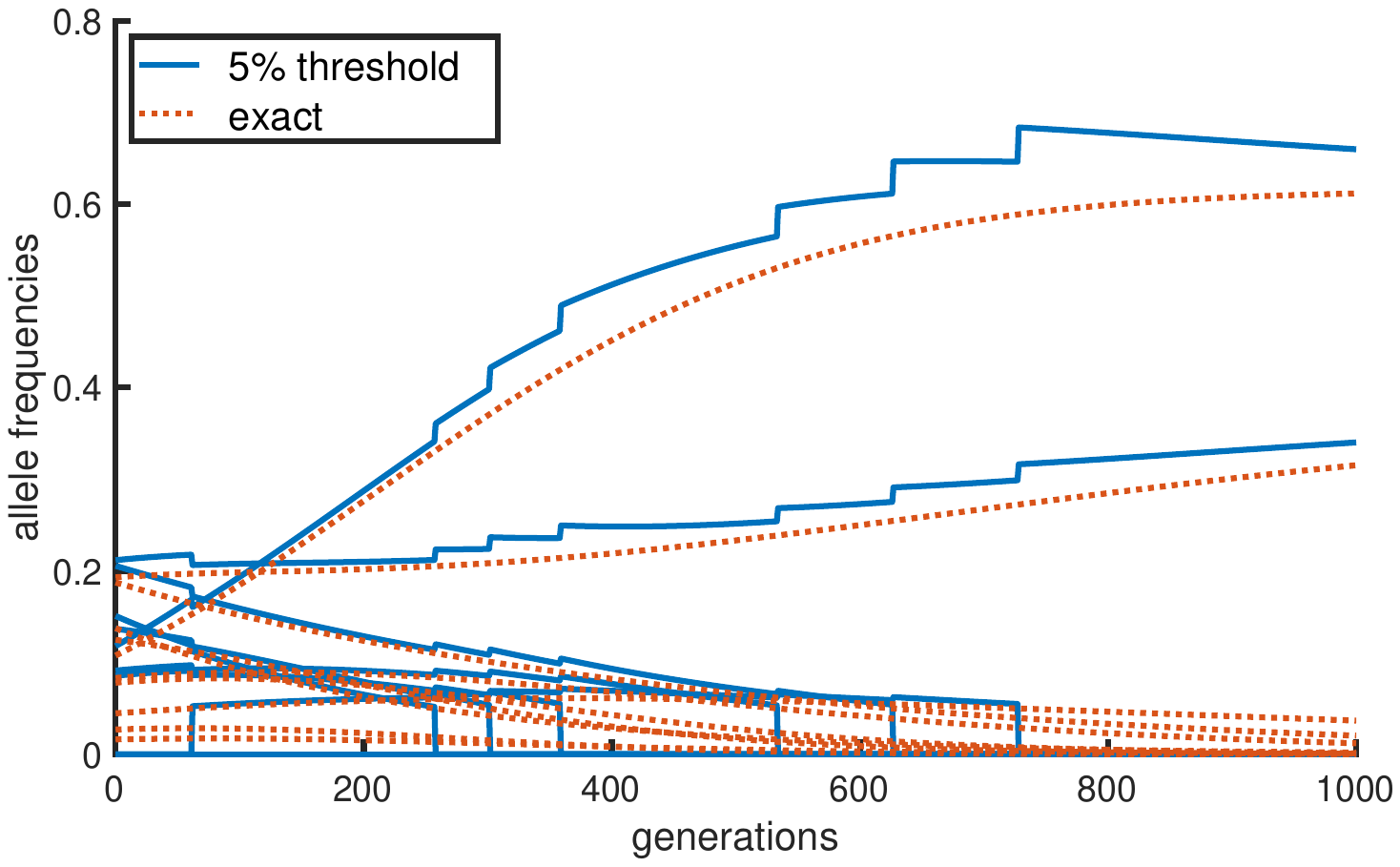} }}
    \caption{\textbf{Illustration of observation thresholds in the analysis of multiallelic trajectories.} (a) Probability to detect a genetic variant at a given frequency assuming different levels of sequencing coverage. This applies to pooled samples. (b) Shown are the exact allele frequency trajectories (red), and those after applying a $5\%$ threshold (blue) for a locus with ten alleles and an arbitrarily chosen scheme of selection. The threshold (i) sets allele frequencies below $5\%$ to zero and (ii) normalizes frequency so that they sum up to unity resulting in the 'steps' in the blue curves.}
    \label{fig:prop_threhold}
\end{figure}

There are a number of practical concerns that we wish to highlight when obtaining data on 
genetic diversity at loci with many alleles. First, obtaining low frequency variants from population genetic data, by pooling data, requires deep sampling to obtain alleles that are segregating at low frequencies. For example, even at 500X coverage the probability of observing a variant that segregates with a frequency of 0.1 percent (e.g. a new mutant in a diploid population of 500 individuals) is less than 50\% (see Figure \ref{fig:prop_threhold}a). So depending on the expected population size it might be necessary to sequence a pooled sample at an enormous coverage. However, sequencing costs have dropped substantially in recent years and therefore deep sequencing may now allow testing of more complex scenarios of selection,
due to the identification of rare genetic variants at multiallelic sites.

Second, incorporating low frequency variants is also crucial when considering time-series experiments since this would allow much improved resolution over time (i.e., avoiding the omission of allele frequency trajectories because they appear hidden at certain time points). This is potentially relevant for evolve and re-sequencing experiments \cite{schlotterer2015} or when tracing strains in metagenomic samples over time \cite{faust2015, hildebrand2019, hildebrand2021}. However, if low-frequency variation is omitted, e.g., as a consequence of a threshold \cite{gossmann2022}, then the allele-frequencies of the remaining (above threshold) alleles might be misleading (see Figure \ref{fig:prop_threhold}b). In time-series experiments this may lead to "jumps" that can be misidentified as signatures of rapid change (e.g. selection). In general, as observation thresholds are to be expected due to sampling effects alone (see Figure \ref{fig:prop_threhold}a) it is evident that much diversity and evolutionary dynamics may be missed because it occurs below the threshold (Figure \ref{fig:prop_threhold}b). 

Third, if SNPs are considered in a population-wide sample, then in principle a single base position can
have four alleles, at most. However, we would expect selection to act at a level of DNA sequences that segregate together and are rarely broken up by recombination events \cite{williams1966}, e.g., at the level of haplotypes \cite{clark2004, fariello2013}, or whole genes \cite{colbran2023}, which will lead to a higher number of alleles. Furthermore, when studying epigenetic variation we can expect an even higher diversity of epialleles because there are more combinations of epigenetic modifications possible than there are unmodified nucleobases \cite{liu2024, kosel2024}. Considering population genetic models with an arbitrary number of alleles is therefore not only realistic, but potentially very important, in order to capture all potential targets of selection \cite{overall2024a}. While it is possible to obtain haplotype information from short read sequencing data \cite{garrison2012}, a better resolution of haplotypes and potential epimarks can be reached when third generation sequencing technologies are applied, such as PacBio \cite{rhoads2015} or Oxford Nanopore \cite{wang2021}. A general approach to obtain complex haplotypes might be to use barcoding approaches from long range sequencing data \cite{sedlazeck2018}.

Fourth, low frequency variation from a population genetic sample might be more difficult to obtain when individuals are sampled, which is not uncommon for vertebrate or other multicellular species. Individual based sampling depth potentially will need to be moderate to identify heterozygous sites, but would then need to be scaled up with the number of individuals tested (see Figure \ref{fig:prop_threhold}a). However, often bioinformatic pipelines take population-wide signatures of polymorphic sites into account to filter out supposed sequencing errors \cite{mckenna2010,korneliussen2014}, which potentially leads to the wrong classification of rare or previously unknown variation. On the other hand, somatic mutations can be mislabeled as germline mutations \cite{dou2018, lupski2013, oota2020}. Because somatic variation does not propagate to subsequent generations, time-series experiments may be used to identify such types of genetic variation, and therefore may help to accurately identify germline mutations segregating at low frequencies.     

\bigskip

\section*{Data availability}
\sloppy

All scripts are available in the github repository \url{https://github.com/NikolasVellnow/selection\_multiallelic} 





\newpage

\bibliographystyle{ieeetr}
\bibliography{Vellnow.bib}


\newpage

\pagebreak
\appendix
\pagestyle{empty}
\begin{center}
    {\Large \textbf{APPENDICES}}
\end{center}

\section{Dynamics of the model}

In this appendix we derive the dynamical equation that the frequencies of
different alleles obey under a Wright-Fisher model. 

We start with an effectively infinite population. 

With $\mathbf{X}(t)$ an $n$ component column vector containing the frequencies
of all $n$ alleles in generation $t$, the processes of reproduction and
selection of the model described in Section 3 of the main text lead to the
frequency obeying the equation
\begin{equation}
\mathbf{X}(t+1)=\mathbf{X}(t)+\mathbf{D}(\mathbf{X}(t)).
\end{equation}
Here $\mathbf{D}(\mathbf{X}(t))$ is an $n$ component column vector. It depends
on the set of frequencies, $\mathbf{X}(t)$, of generation $t$ and contains the
changes of the frequencies that occur, due to \textit{natural selection} in
generation $t$. In Appendix B we give details of the form of $\mathbf{D}
(\mathbf{x})$, which we interpret as the selective evolutionary force that
acts when the set of allele frequencies is $\mathbf{x}$.

In a finite population, apart from the occurrence of reproduction and
selection, the process of thinning of the population occurs. This amounts to
\textit{non selective sampling} that leads to $N$ adults being present at the start of
each generation. Under the appropriate generalisation of the Wright-Fisher
model \cite{nagylaki1992}, this has the result that the set of allele frequencies in generation
$t+1$, namely $\mathbf{X}(t+1)$, does not, generally, coincide with $\mathbf{X}
(t)+\mathbf{D}(\mathbf{X}(t))$. Rather, $\mathbf{X}(t+1)$ is
\textit{distributed around} the value $\mathbf{X}(t)+\mathbf{D}(\mathbf{X}
(t))$ according to
\begin{equation}
2N\mathbf{X}(t+1)\sim\operatorname*{Mult}(2N,\mathbf{X}(t)+\mathbf{D}
(\mathbf{X}(t)).\label{dist eq}
\end{equation}
Here $\operatorname*{Mult}(m,\mathbf{p})$ denotes a \textit{multinomial distribution},
where the parameter $m$ is the number of independent trials, while 
the parameter $\mathbf{p}$ is a column vector containing the probabilities of falling into 
different categories, on each trial\footnote{We note that Eq. (\ref{dist eq}) is an 
approximation when fitnesses are not multiplicative, however for $|A_{i,j}| \ll 1$, as we shall assume,
it is a good approximation \cite{nagylaki1992} (P. 252).}.

Let $\mathbf{M}(t)$ denote a multinomial random variable with distribution
$\operatorname*{Mult}(2N,\mathbf{X}(t)+\mathbf{D}(\mathbf{X}(t))$. The form of
$\mathbf{M}(t)$ is an $n$ component column vector that has a conditional
expectation of
\begin{equation}
E\left[  \mathbf{M}(t)|\mathbf{X}(t)=\mathbf{x}\right]  =2N\left[
\mathbf{x}+\mathbf{D}(\mathbf{x})\right].
\end{equation}
It follows that equivalent to Eq. (\ref{dist eq}) is the equation $\mathbf{X}
(t+1)=\mathbf{M}(t)/(2N)$ which we write as
\begin{equation}
\mathbf{X}(t+1)=\mathbf{X}(t)+\mathbf{D}(\mathbf{X}(t))+\boldsymbol{\xi}(t)
\end{equation}
where
\begin{equation}
\boldsymbol{\xi}(t)=\frac{\mathbf{M}(t)}{2N}-\left[  \mathbf{X}(t)+\mathbf{D}
(\mathbf{X}(t))\right].
\end{equation}
The quantity $\boldsymbol{\xi}(t)$ represents statistical fluctuations, due to
non selective sampling, around the `deterministic prediction' of
$\mathbf{X}(t+1)$, namely $\mathbf{X}(t)+\mathbf{D}(\mathbf{X}(t))$, and $\boldsymbol{\xi}(t)$ can be
interpreted as a stochastic `force' on allele frequencies that arises from
random genetic drift.

From properties of the multinomial distribution \cite{evans1993}, it follows that the
conditional expectation of $\boldsymbol{\xi}(t)$ vanishes, i.e.,
\begin{equation}
E\left[  \boldsymbol{\xi}(t)|\mathbf{X}(t)=\mathbf{x}\right]  =\mathbf{0}
\label{Exi=0}
\end{equation}
(the right hand side is an $n$ component column vector of zeros).
Additionally, the variance-covariance matrix of $\boldsymbol{\xi}(t)$ is
\begin{equation}
E\left[  \boldsymbol{\xi}(t)\boldsymbol{\xi}^{T}(t)|\mathbf{X}(t)=\mathbf{x}
\right]  =\frac{\mathbf{V}(\mathbf{x}+\mathbf{D}(\mathbf{x}))}{2N}
\end{equation}
where $\mathbf{V}(\mathbf{x})$ is the matrix
\begin{equation}
\mathbf{V}(\mathbf{x})=\operatorname*{diag}(\mathbf{x})-\boldsymbol{xx}^{T}.
\end{equation}
whose elements are given by $V_{i,j}(\mathbf{x})=x_i\delta_{i,j}-x_ix_j$.

On the assumption that all elements of $\mathbf{D}(\mathbf{x})$ are
small compared with the corresponding element of $\mathbf{x}$, which amounts
to saying selection causes a very small \textit{fractional} change in allele
frequencies, as commonly holds, we have the standard approximation \cite{ewens2004}
\begin{equation}
E\left[  \boldsymbol{\xi}(t)\boldsymbol{\xi}^{T}(t)|\mathbf{X}(t)=\mathbf{x}
\right]  \simeq\frac{\mathbf{V}(\mathbf{x})}{2N}.\label{var cov xi =V/2N}
\end{equation}
We thus arrive at a description of the dynamics given by $\mathbf{X}
(t+1)=\mathbf{X}(t)+\mathbf{D}(\mathbf{X}(t))+\boldsymbol{\xi}(t)$ with the
primary properties of $\boldsymbol{\xi}(t)$ given by Eqs. (\ref{Exi=0}) and
(\ref{var cov xi =V/2N}).

\section{Deriving the form of the selective force}

In this appendix we derive a compact representation of the evolutionary force
associated with selection. 

We have $n$ alleles labelled $B_{i}$ with $i=1,2,\ldots,n$ and assume the
$B_{i}B_{j}$ genotype has a fitness proportional to $1+A_{i,j}$ with
$A_{i,j}=A_{j,i}$.

Let $x_{i}$  be the frequency of allele $B_{i}$ in the adults of a particular
generation, and let $x_{i}^{\prime}$ be the corresponding frequency in
offspring, after random mating, followed by selection, has occurred. 

Contributions to $x_{i}^{\prime}$ arise from: (i) offspring with the
$B_{i}B_{i}$ genotype and (ii) offspring with the $B_{i}B_{j}$ genotype, with
$j\neq i$. With all sums running from $1$ to $n$ unless otherwise stated, this
leads to
\begin{equation}
x_{i}^{\prime}=\tfrac{2\left(  1+A_{i,i}\right)  x_{i}^{2}+\sum_{j(\neq
i)}\left(  1+A_{i,j}\right)  2x_{i}x_{j}}{2\left[  \left(  1+A_{i,i}\right)
x_{i}^{2}+\sum_{j(\neq i)}\left(  1+A_{i,j}\right)  2x_{i}x_{j}+\sum_{j(\neq
i),k(\neq i)}\left(  1+A_{j,k}\right)  x_{j}x_{k}\right]  }.
\end{equation}
This can be simplified to
\begin{equation}
x_{i}^{\prime}=\frac{x_{i}\left(  1+\sum_{j}A_{i,j}x_{j}\right)  }
{1+\sum_{j,k}A_{j,k}x_{j}x_{k}}.
\end{equation}
Alternatively, we have
\begin{equation}
x_{i}^{\prime}=x_{i}+\frac{x_{i}\left(  \sum_{j}A_{i,j}x_{j}-\sum_{j,k}
A_{j,k}x_{j}x_{k}\right)  }{1+\sum_{j,k}A_{j,k}x_{j}x_{k}}\label{xi'-xi}
\end{equation}
and we rewrite this equation as
\begin{align}
x_{i}^{\prime}  & =x_{i}+\frac{\sum_{j,k}\delta_{i,j}x_{j}A_{j,k}x_{k}
-\sum_{j,k}x_{i}x_{j}A_{j,k}x_{k}}{1+\sum_{j,k}x_{j}A_{j,k}x_{k}}\nonumber\\
& \nonumber\\
& =x_{i}+\frac{\sum_{j,k}\left(  \delta_{i,j}x_{j}-x_{i}x_{j}\right)
A_{j,k}x_{k}}{1+\sum_{j,k}x_{j}A_{j,k}x_{k}}.\label{xi'-xi 2}
\end{align}
Using $[\mathbf{y}]_{i}$ to denote the $i$'th element of the vector
$\mathbf{y}$, and with $\mathbf{V}(\mathbf{x})$ the $n\times n$ matrix with
elements
\begin{equation}
V_{i,j}(\mathbf{x})=x_{i}\delta_{i,j}-x_{i}x_{j}\label{V}
\end{equation}
we can write Eq. (\ref{xi'-xi 2}) as $x_{i}^{\prime}=x_{i}+\frac{\left[
\mathbf{V}(\mathbf{x})\mathbf{Ax}\right]  _{i}}{1+\mathbf{x}^{T}\mathbf{Ax}}$
where $\mathbf{A}$ is a matrix with elements $A_{i,j}$. From this we obtain
\begin{equation}
\mathbf{x}^{\prime}=\mathbf{x}+\frac{\mathbf{V}(\mathbf{x})\mathbf{Ax}
}{1+\mathbf{x}^{T}\mathbf{Ax}}
\end{equation}
and this equation allows us to identify the second term on the right hand side with
the evolutionary force of selection, i.e., we identify
\begin{equation}
\mathbf{D}(\mathbf{x})=\frac{\mathbf{V}(\mathbf{x})\mathbf{Ax}}{1+\mathbf{x}
^{T}\mathbf{Ax}}.
\end{equation}

\end{document}